\documentclass[amsmath,amssymb,aps,prb,twocolumn,superscriptaddress,longbibliography]{revtex4-2}
\usepackage{graphicx}
\usepackage{dcolumn}
\usepackage{color}
\usepackage{hyperref}
\usepackage{booktabs}
\usepackage{physics}
\usepackage{multirow}

\renewcommand{\cite}[1]{[\onlinecite{#1}]}

\newcommand{\eps}{\varepsilon}      
\newcommand{\bas}[1]{{\bf e}_{#1}}
\renewcommand{\vec}[1]{{\bf #1}}

\begin{document}

\title{Three-dimensional higher-order topological insulator protected by cubic symmetry}

\author{Valerii~I.~Kachin}
\affiliation{School of Physics and Engineering, ITMO University, Saint Petersburg 197101, Russia}

\author{Maxim~A.~Gorlach}
\affiliation{School of Physics and Engineering, ITMO University, Saint Petersburg 197101, Russia}
\email{m.gorlach@metalab.ifmo.ru}


\begin{abstract}
Recently discovered photonic higher-order topological insulators enable unprecedented flexibility in the robust localization of light in structures of different dimensionality. While the potential of the two-dimensional systems is currently under active investigation, only a few studies explore the physics of the three-dimensional higher-order topological insulators. Here we propose a three-dimensional structure with cubic symmetry exhibiting vanishing bulk polarization but nonzero corner charge and hosting a zero-dimensional corner state mediated by the long-range interactions. We trace the evolution of the corner state with the next-nearest-neighbor coupling strength and prove the topological origin of the corner mode calculating the associated topological invariants. Our results thus reveal the potential of long-range couplings for the formation of three-dimensional higher-order topological phases.
\end{abstract}

\maketitle

\section{INTRODUCTION}

Topological photonics offers the control over light propagation and localization by engineering the topology of the bands in the reciprocal space complementing the conventional dispersion engineering~\cite{Ozawa_RMP}. Typically, the topological states have the dimensionality lower by one than the dimensionality of the structure. This limitation, however, has been overcome with the discovery of higher-order topological insulators   hosting the topological states with lower dimensionalities~\cite{Benalcazar2017, Schindler2018}.

So far, active research of higher-order topological structures was mainly focused on the two-dimensional (2D) systems realized at such platforms as resonant electric circuits~\cite{Imhof, Serra-Garcia-Huber,Nejad2019}, acoustic resonators~\cite{Xue2018,Ni2018}, microwave~\cite{Benalcazar2018, Khanikaev-Zhirihin2020} and optical setups~\cite{Mittal-NP, Ashraf2019}.

Three-dimensional topological phases remain much less explored with only few first experimental studies currently available~\cite{Bao2019,Weiner2020,Liu-Ma-Zhang2020}. At the same time, such systems are expected to enable rich physics due to the anticipated hierarchy of two-dimensional (2D), one-dimensional (1D) and zero-dimensional (0D) topological states coexisting in the same structure~\cite{Nag2021,Mook2020} and serving as topological waveguides and topological resonators for transmission and storage of light.

The physics of the three-dimensional topological phases is typically understood based on the tight-binding model which includes only the coupling of the nearest neighbors. On the other hand, long-range nature of electromagnetic interactions may facilitate the topological states beyond the tight-binding description, as has been recently demonstrated for 2D structures~\cite{Khanikaev-Zhirihin2020,Jiang2020,Olekhno2021}. At the same time, the impact of long-range interactions on three-dimensional topological phases remains vastly unexplored.

To fill this gap and make a step towards future technologies of compact topological resonators, here we put forward and investigate a three-dimensional cubic structure with $O_h$ symmetry group hosting a  zero-dimensional corner-localized topological state. As we prove, long-range interactions play the key role in the opening of a complete bandgap around corner mode energy, while the magnitude of the next-nearest-neighbor coupling constants directly controls the localization length of the topological corner state as well as its spectral separation from the continuum of bulk and surface states.

The rest of the article is organized as follows. In Section~\ref{sec:design}, we overview the proposed three-dimensional structure, construct its Bloch Hamiltonian and explore the dispersion of the bulk bands. Exploiting the calculated bulk modes, in Sec.~\ref{sec:inv} we examine the symmetries of the model and develop the procedure to calculate the topological invariants in our 3D case including bulk polarization and corner charge. Section~\ref{sec:finite} continues with the analysis of the states supported by the finite 3D system assessing their localization properties and the conditions for the existence of in-gap corner-localized state. In the concluding Sec.~\ref{sec:con}, we discuss the obtained results highlighting possible experimental implementations, further perspectives and potential applications. Technical details of our calculations are summarized in Appendixes~\ref{supp:a}-\ref{supp:e}.

\section{Bloch Hamiltonian and bulk modes}\label{sec:design}

\begin{figure}[!ht]
\center{\includegraphics[scale=0.5]{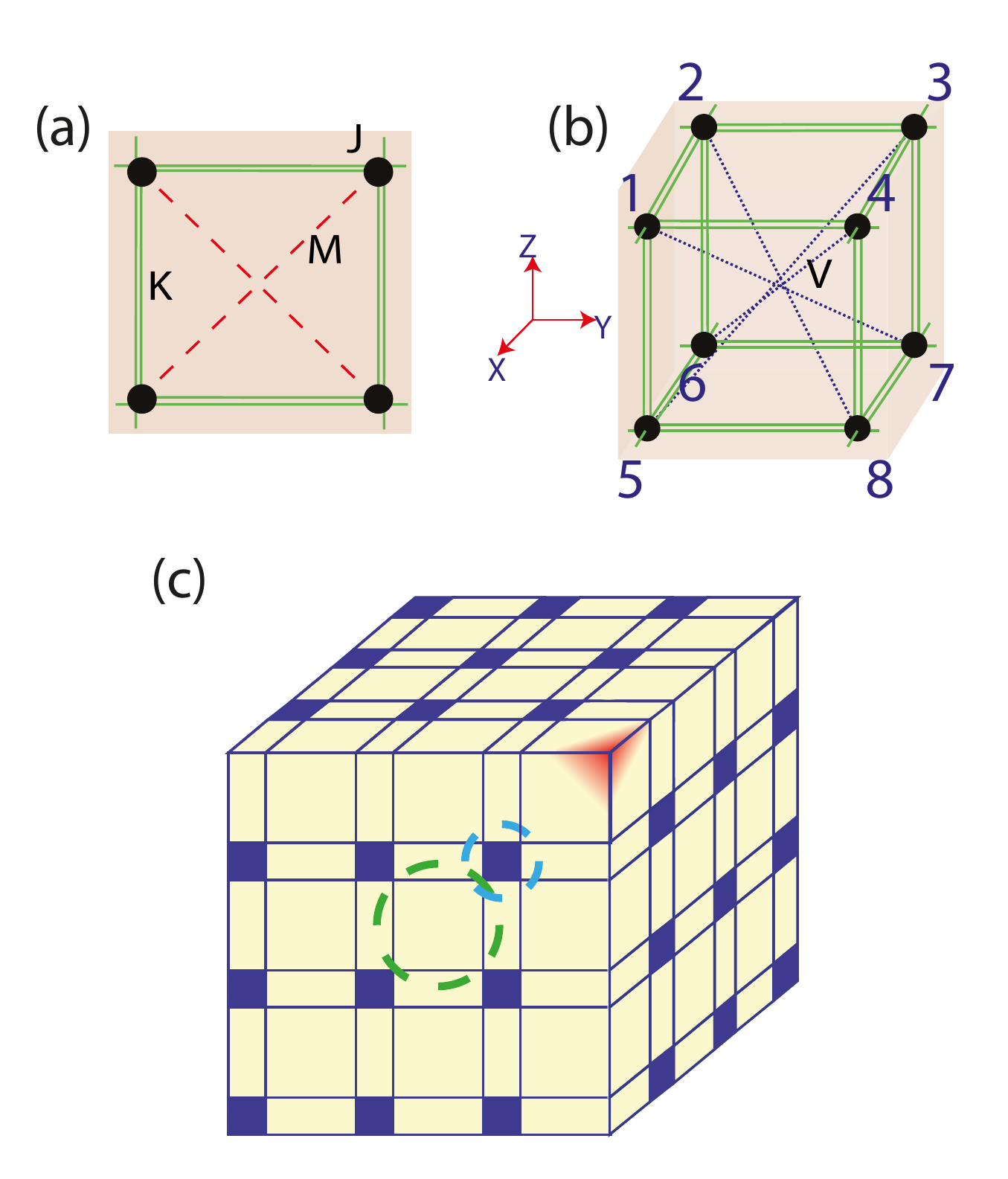}}
\caption{Geometry of the proposed three-dimensional higher-order topological insulator. Nearest-neighbor couplings $J$ and $K$, both positive, are shown by the single and double green lines, respectively. Two additional types of the next-nearest neighbor couplings with the amplitudes $M$ and $V$ are shown by the red dashed and blue dotted lines, respectively. (a) Single face of the unit cell. (b) Three-dimensional unit cell and enumeration of the sites used to construct Bloch Hamiltonian Eq.~\eqref{eq:Hamiltonian}. (c) A finite sample of the three-dimensional structure. Dashed circles indicate two possible choices of the unit cell with strong and weak links inside. A corner highlighted by red hosts the topological corner state.}\label{fig:str}
\end{figure}

A single face and the unit cell of the proposed periodic structure are depicted in Fig.~\ref{fig:str}(a,b). Note that contrary to the canonical cases of quadrupole and octupole insulators~\cite{Benalcazar2017} the couplings of the nearest neighbors within the unit cell are positive and identical. Hence the periodic structure possesses $O_h$ symmetry group.

Furthermore, the alternating pattern of the coupling links along each of the coordinate axes resembles  well-celebrated Su-Schrieffer-Heeger model (SSH)~\cite{Su} providing its three-dimensional generalization illustrated in Fig.~\ref{fig:str}(c).

Assuming for the moment only the interaction of the nearest neighbors, we arrive to the Bloch Hamiltonian

\begin{equation}\label{eq:Hamiltonian}
\begin{gathered}
    H(k_x,k_y,k_z)=\left(
        \begin{tabular}{c|c}
          $H_0(k_x,k_y)$   & $H_1(k_z)$  \\
          \hline
          $H_1^\dag(k_z)$   & $H_0(k_x,k_y)$
        \end{tabular}
    \right),
\end{gathered}
\end{equation}
where $H_0(k_x,k_y)$ is a $4\times 4$ block that corresponds to the Hamiltonian of a single face in $Oxy$ plane detached from the rest of the structure, i.e. two-dimensional SSH 
with intra- and intercell coupling constants $K$ and $J$, respectively:
\begin{small}
\begin{equation}
\label{eq:ham1}
    H_0 = 
    \left( \begin{tabular}{c c c c}
        $0$ & $K + J e^{i k_x}$ &$ 0$ & $K + J e^{-i k_y}$\\ 
        $K + J e^{-i k_x}$ & $0$ & $K + J e^{-i k_y}$ & $0$ \\ 
        $0$ & $K + J e^{i k_y}$ & $0$ & $K + J e^{-i k_x}$ \\
        $K + J e^{i k_y}$ & $0$ & $K + J e^{i k_x}$ & $0$
    \end{tabular}\right).
\end{equation}
\end{small}
Off-diagonal block $H_1(k_z)$ describing the coupling of the individual layers along $z$ axis takes the form:
\begin{equation}
    H_1=\left(K + J e^{i k_z}\right)\,\hat{I}_4\:,
\end{equation}
where $\hat{I}_4$ is a $4\times 4$ identity matrix.

The described nearest-neighbor-coupled structure does not support spectrally isolated corner states, and the spectrum appears to be gapless at zero energy. However, the interaction of the distant neighbors can dramatically alter the situation.

First, we incorporate next-nearest neighbor interaction along the diagonals of the faces described by the matrix $\widehat{M}$ such that the Hamiltonian of a single face $H_0(k_x,k_y)$ acquires an additional contribution $\widehat{M}$, where 
\begin{equation}
\label{eq:MOp}
    \widehat{M}= M
    \left( \begin{tabular}{c c c c}
        $0$ & $0$ & $1$ & $0$\\ 
        $0$ & $0$ & $0$ & $1$\\ 
        $1$ & $0$ & $0$ & $0$\\
        $0$ & $1$ & $0$ & $0$
    \end{tabular}\right).
\end{equation}
The coupling links described by the operator $\widehat{M}$ are shown in Fig.~\ref{fig:str}(a) by red dashed lines.

Even though this type of interaction opens the topological gap in 2D  case~\cite{Olekhno2021}, it appears to be insufficient to open a complete bandgap in the 3D scenario. Therefore, we introduce additionally one more type of long-range interaction along the principal diagonal of the cube with the amplitude $V$ illustrated in Fig.~\ref{fig:str}(c) by blue dotted lines. 

These two types of the long-range coupling yield the following form of the Bloch Hamiltonian:
\begin{equation}\label{eq:HamiltonianLR}
\begin{gathered}
    \widehat{H}(k_x,k_y,k_z)=\left(
        \begin{tabular}{c|c}
          $H_0(k_x,k_y)+\widehat{M}$   & $H_1(k_z) +\widehat{L}$ \\
          \hline
          $H_1^\dag(k_z)+\widehat{L}$   & $H_0(k_x,k_y)+\widehat{M}$
        \end{tabular}
    \right),
\end{gathered}
\end{equation}
where 
\begin{equation}
\label{eq:LR2}
    \widehat{L}= 
    \left( \begin{tabular}{c c c c}
        $0$ & $M$ & $V$ & $M$\\ 
        $M$ & $0$ & $M$ & $V$\\ 
        $V$ & $M$ & $0$ & $M$\\
        $M$ & $V$ & $M$ & $0$
    \end{tabular}\right).
\end{equation}

\begin{figure}[!ht]
\center{\includegraphics[scale=1]{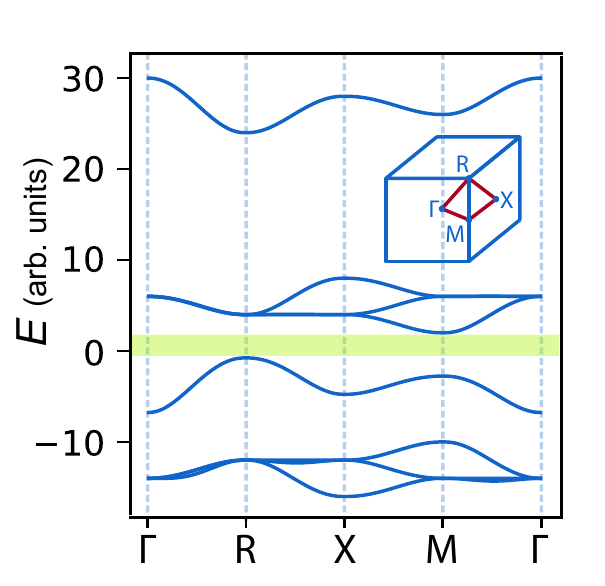}}
\caption{Calculated band structure of the proposed model with nearest-neighbor couplings $J = 1$, $K = 6$ and long-range interactions $M = 4$ and $V = -3$. The complete bandgap which opens at energies close to zero is shaded by green. Inset shows the first Brillouin zone and the associated high-symmetry points.}\label{fig:bandstr}
\end{figure}

Inspecting the band structure of the constructed Hamiltonian (Fig.~\ref{fig:bandstr}), we observe the opening of a complete bandgap at energies close to zero.

\section{Generalized chiral symmetry and topological invariant}\label{sec:inv}

Besides opening of a complete bandgap at energies close to zero, the proposed system features the so-called generalized chiral symmetry. This symmetry implies the existence of an operator $\widehat{\Gamma}_8$ such that $\widehat{\Gamma}_8^8=\hat{I}_8$ ($\hat{I}_8$ is a unity $8\times 8$ matrix) which generates a sequence of the matrices $H_n=\widehat{\Gamma}_8^n H \widehat{\Gamma}_8^{-n}$ that satisfy the condition
\begin{equation}\label{eq:genchs}
\sum_{n=1}^7 H_n+H=0.    
\end{equation}

In the other words, states from the eight bulk bands are separated into the groups, each contains 8 states. Sum of energies of the modes comprising the same group is equal to zero, while the wave functions of these modes are mutually related via generalized chiral symmetry operator $\widehat{\Gamma}_8$.

We construct $\widehat{\Gamma}_8$ as a diagonal matrix with the entries at the principal diagonal equal to the roots of the equation $x^8=1$, i.e. $\left(\Gamma_8\right)_{n,n}=\exp\left(\frac{i(n-1)\pi}{4}\right)$. 

It should be emphasized that the similar type of generalized chiral symmetry occurs in the  three-dimensional models based on pyrochlore lattice~\cite{Weiner2020} as well as two-dimensional structures with $D_3$ and $D_4$ symmetry groups~\cite{Ni2018,Olekhno2021}.

Intrinsic symmetries of the model point towards its possible topological origin. To verify this, we extend the approach of Ref.~\cite{Benalcazar2019} to the 3D case examining the behavior of the bulk modes in the high-symmetry points of the Brillouin zone under symmetry transformations. From the obtained symmetry indices, we evaluate bulk polarization and extract the corner charge.

The high-symmetry points (also termed Wyckoff positions) are chosen such that they remain invariant or transit to the equivalent points under the symmetry operations of $O_h$ group. These points include the center of the Brillouin zone $\Gamma$ with ${\bf k}=(0,0,0)$; the center of the cubic face $X$ with ${\bf k}=(\pi,0,0)$; the center of the cube edge $M$ with ${\bf k}=(\pi,\pi,0)$ and cube vertex $R$ with ${\bf k}=(\pi,\pi,\pi)$. Due to symmetry, the Brillouin zone contains several $X$, $M$ and $R$ points connected to each other via symmetry transformations.

In turn, the entire symmetry group $O_h$ can be generated by the four principal symmetry elements which include inversion $i$ and the three types of rotation: $C_2$, $C_3$ and $C_4$~\cite{Dresselhaus}. These elements commute with the Bloch Hamiltonian in high-symmetry points which allows one to label the eigenstates by the respective eigenvalues of the symmetry operators.  In this approach, the change in the number of the  bulk bands with a given symmetry index below or above the bandgap signals the topological nature of the model.

Based on these considerations, we define the topological invariant for $O_h$-symmetric model as a set of the following numbers:
\begin{equation}
\begin{gathered}
     \chi = (\#X_1^{(i)}-\#\Gamma_1^{(i)}, \#M_1^{(C_2)}-\#\Gamma_1^{(C_2)}, \\   \#R_1^{(C_3)}-\#\Gamma_1^{(C_3)}, \#R_1^{(C_4)}-\#\Gamma_1^{(C_4)}),
\label{eq:chi}
\end{gathered}
\end{equation}
where the upper index denotes the applied symmetry operator, lower index is associated with the behavior of the mode under the symmetry transformation, and $\#$ indicates the number of the eigenstates with a given transformation law below the bandgap in $\Gamma$, $R$, $M$ or $X$ points of the first Brillouin zone.

While the detailed calculation of the topological invariant is provided in Appendix~\ref{supp:a}, here we discuss the main results. If the choice of the unit cell is consistent with that used previously [Fig.~\ref{fig:str}(b), see also the blue circle in Fig.~\ref{fig:str}(c)], the topological invariant is equal to $\chi=(0,0,0,0)$ indicating that the topological corner states are absent at the strong link corner. However, for another choice of the unit cell with the weak links inside [Fig.~\ref{fig:str}(c), green circle], the topological invariant takes the value $\chi=(-2,0,0,2)$ predicting the topological corner state at the weak link corner.

Using the obtained symmetry indices, we evaluate bulk polarization for the bands below  zero-energy bandgap. As demonstrated in Appendix~\ref{supp:b}, bulk polarization $P_x=P_y=P_z$ is related to the inversion eigenvalue in $X$ point as
\begin{equation}\label{eq:BulkPolariz}
    P_x=\frac{1}{2}\,\left[X_2^{J}\right]\:,
\end{equation}
where $[X_2^J]=\#X_2^J-\#\Gamma_2^J$, and $\#X_2^J$ ($\#\Gamma_2^J$) is the number of  eigenstates odd under inversion below the bandgap in $X$ and $\Gamma$ points, respectively. Note that Eq.~\eqref{eq:BulkPolariz} provides a natural generalization of the two-dimensional result~\cite{Benalcazar2019}.

Since $[X_2^J]$ is even for both unit cell choices, bulk polarization vanishes in both cases. However, vanishing bulk polarization does not mean that the model is topologically trivial. Corner charge still can be nonzero, as is the case for several 2D systems~\cite{Benalcazar2019}. To check this, we choose the unit cell with the weak links inside and compute bulk polarization for each of the bands below the bandgap separately via Eq.~\eqref{eq:BulkPolariz}. We recover that the two bands have polarization $1/2$, while the remaining two bands have zero polarization.

As further discussed in Appendix~\ref{supp:c}, this indicates that two Wannier centers occupy the center of the unit cell, while the remaining two are located at the corner of the unit cell. The latter Wannier centers define the corner charge of the system equal to 1/4 for all of the corners. The derived quantization of the corner charge thus proves the topological nature of the studied system.

It should be also stressed that even though the next-nearest neighbor interactions $M$ and $V$ are crucial for opening of a complete bandgap at energies close to zero, they do not affect the value of the topological invariant. Together with generalized chiral symmetry of the model this means that for the certain parameter values corner state exists in the continuum of the bulk modes providing the realization of bound state in the continuum~\cite{SoljacicBIC}.

\section{Modes of a finite system}\label{sec:finite}

As a next step, we examine the modes of a finite 3D structure depicted in Fig.~\ref{fig:str}(c). To probe two possible terminations of the structure simultaneously, we consider a cube with a half-integer number of unit cells along each edge which renders the symmetry of the finite sample lower than that of a periodic structure.

To quantify mode localization properties, we calculate an inverse participation ratio ($IPR$) defined as~\cite{THOULESS197493}
\begin{equation}
    IPR=\left(\sum\limits_{m,n,l}\,|\psi_{mnl}|^4\right)\,\left(\sum\limits_{m,n,l}\,|\psi_{mnl}|^2\right)^{-2}\:,
\end{equation}
where $\psi_{mnl}$ are the amplitudes of a given mode in the sites of the 3D system (field strength, voltage, pressure, etc. depending on the chosen physical realization) and the summation is performed over all sites of the lattice. 

\begin{figure}[ht!]
\center{\includegraphics[scale=0.7]{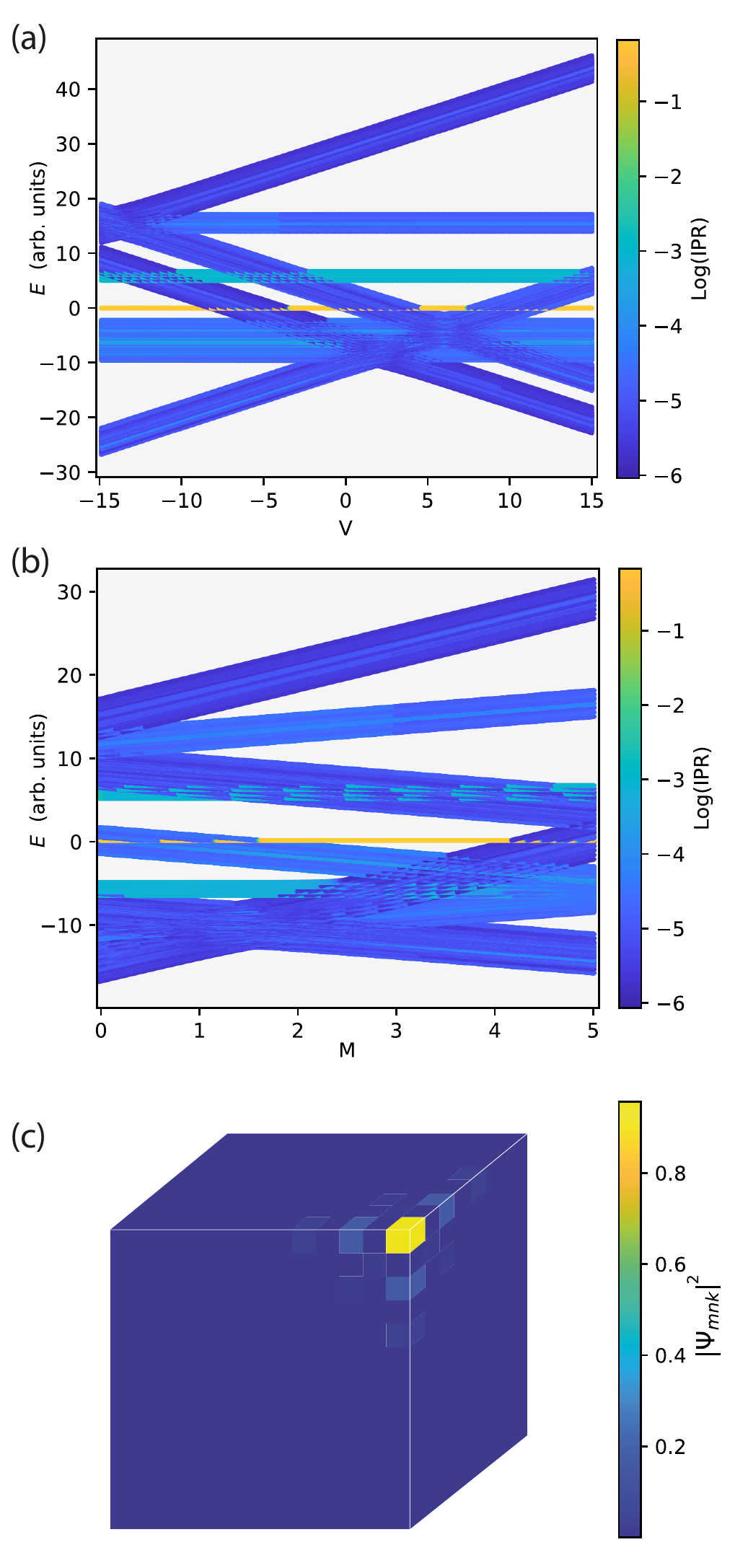}}
\caption{Eigenmodes of a finite 3D structure. (a) Mode energies for $9\times 9\times 9$ structure with coupling constants $J = 1$, $K = 6$, $M = 2.7$ versus long-range coupling $V$, $-15 \leq V \leq 15$. (b) Mode energies for $9\times 9\times 9$ structure with coupling constants $J = 1$, $K = 6$, $V = -3$ versus long-range coupling $M$, $0 \leq M \leq 5$. Color encodes the logarithm of the inverse participation ratio (IPR) for the eigenmodes. (c) Eigenmode profile corresponding to the in-gap corner state in $13 \times 13 \times 13$ structure with coupling constants $J = 1$, $K = 6$, $M = 4$, $V = -3$. Color shows the absolute value of the mode amplitude at a given site of the lattice.}\label{fig:finres}
\end{figure}

A finite structure supports four types of modes: bulk states spread over the entire sample; surface states localized at the faces of the cubic structure; hinge states propagating along the edges of the cube; corner states pinned to the cube vertices. Accordingly, the inverse participation ratio for these states exhibits four different types of scaling with the system size $N$: $\propto 1/N^3$ for bulk states; $\propto 1/N^2$ for surface states; $\propto 1/N$ for hinge states and a value close to 1 for the corner mode.

In Fig.~\ref{fig:finres}(a), we trace the evolution of the modes supported by the finite 3D structure as a function of the long-range coupling $V$ keeping the values of the other coupling constants fixed. Using the color to encode the localization properties of the modes, we observe four distinct colors at the diagram corresponding to the expected four types of modes. 

Dark blue color corresponds to the bulk modes occupying the entire lattice. By their nature, such modes are sensitive to the long-range coupling exhibiting a pronounced dependence of energy on the magnitude of $V$.

\begin{figure}[ht!]
\center{\includegraphics[scale=0.57]{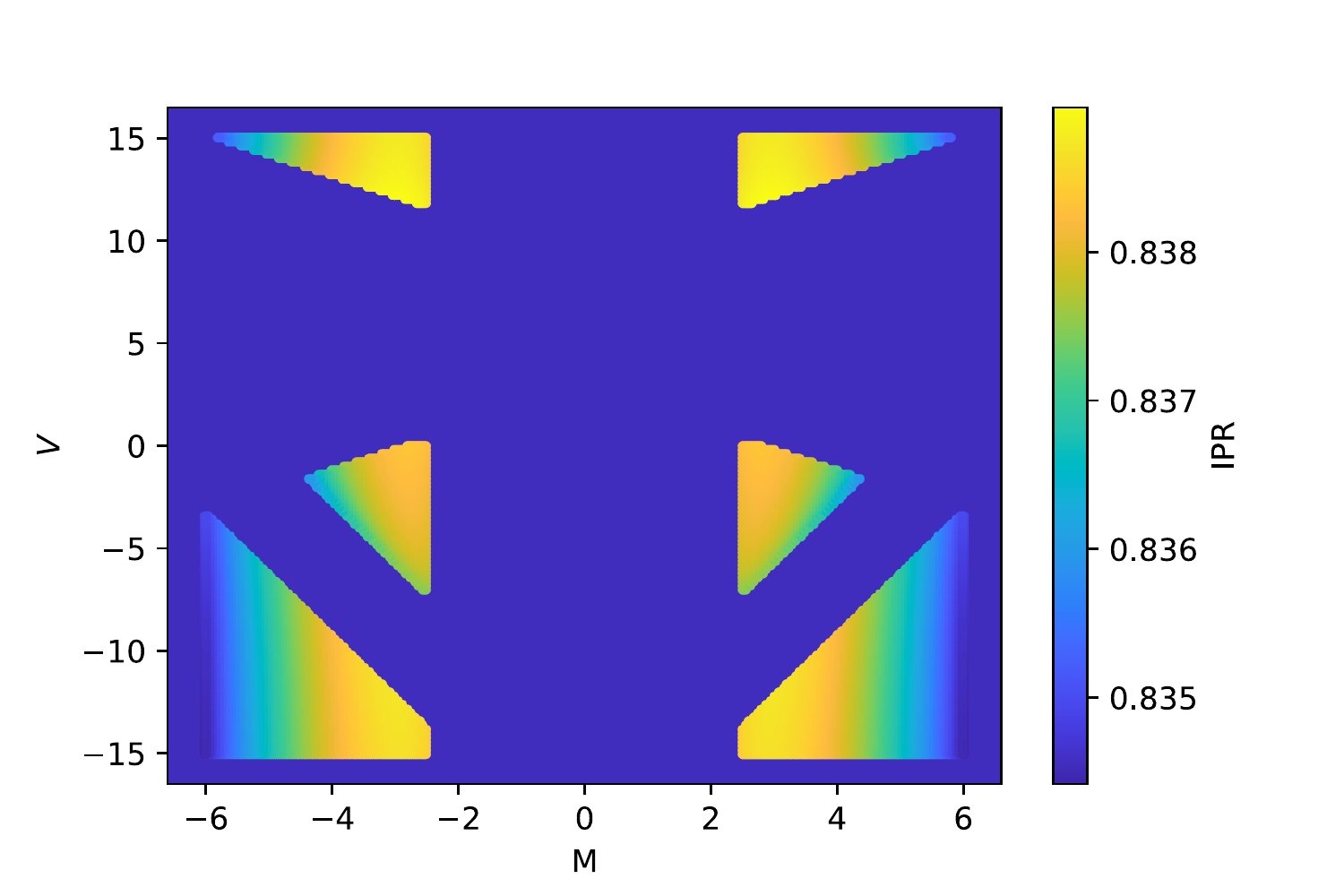}}
\caption{Colormap showing the inverse participation ratio ($IPR$) of the corner mode as a function of the next-nearest neighbor couplings $M$ and $V$ calculated for $9 \times 9 \times 9$ system with $J = 1$ and $K = 6$, when the corner state exists inside the band gap.}\label{fig:mvsc}
\end{figure}

At the same time, modes shown by light blue and teal colors are almost unaffected by $V$. Indeed, since these states are mostly localized at the faces and edges of the cube, respectively, the long-range coupling has no effect on their energies. The difference between them, however, becomes evident via their dependence on $M$ coupling amplitude [Fig.~\ref{fig:finres}(b)]: while energies of the surface modes exhibit characteristic dependence on $M$, energies of the hinge states remain almost unaffected.

Finally, we observe a single state with the highest $IPR$ that corresponds to the corner-localized mode. Depending on parameters, such state can either overlap with the continuum of bulk modes or emerge inside the bandgap. The characteristic profile of this mode shown in Fig.~\ref{fig:finres}(c) resembles that in the canonical Su-Schrieffer-Heeger model by the staggered pattern of amplitudes. The energy of the state, however, is different from zero and the localization is not captured by the simple exponential formula.

Furthermore, the localization of the corner mode is strongly affected by the long-range coupling parameters $M$ and $V$ as illustrated by the phase diagram Fig.~\ref{fig:mvsc}. Specifically, even small values of long-range coupling $V$ may enable relatively well-localized states, while coupling $M$ should be sufficiently strong exceeding the threshold value $|M|>2$.

Even if the corner mode coexists with the continuum of the bulk states, the topological origin of the model is conserved. In such case, however, the outlined procedure to calculate the topological invariant is no longer valid. Instead, one has to examine local density of states associated with the corner of the 3D structure following the approach of Ref.~\cite{Peterson1114}. As further discussed in Appendix~\ref{supp:d}, a pronounced peak in the frequency dependence of the local density of states can be observed even in the regime of corner state in the continuum, proving the existence of the localized mode and heralding the topological nature of the studied system. 

Nontrivial topological properties of our structure are manifested not only via corner states, but also via interface states which appear if two cubic samples with the opposite dimerizations are stacked [Fig.~\ref{afig:inter}(a)]. In such case, we observe a zero-dimensional localized mode with the energy close to that of the corner state [Fig.~\ref{afig:inter}(b-d)].

At the same time, the corner state is quite robust to disorder in the nearest-neighbor couplings and, partially, even to disorder in the long-range interactions as further analyzed in Appendix~\ref{supp:e} (see Fig.~\ref{afig:robust}).

\section{DISCUSSION AND CONCLUSIONS}
\label{sec:con}

In conclusion, we have put forward a three-dimensional higher-order topological insulator that supports an in-gap topological corner state protected by $O_h$ lattice symmetry and featuring a quantized corner charge. While the model with the nearest-neighbor coupling is gapless at zero energy supporting the corner mode in the continuum of the bulk states,  long-range interactions introduced in our system open a complete bandgap at energies close to zero and enable the topological corner state inside this gap.

It should be emphasized that the proposed system can be readily realized experimentally with the help of three-dimensional acoustic metamaterials. Recent experiments have demonstrated topological corner states in a three-dimensional breathing pyrochlore lattice~\cite{Weiner2020}, three-dimensional Su-Schrieffer-Heeger model~\cite{Zheng2020} as well as octupole insulators~\cite{Ni2020,Xue2020}. Note that the latter structures require the combination of positive and negative couplings which can be flexibly engineered in  acoustic setups~\cite{Ni2020,Xue2020}. Furthermore, recent works~\cite{He2020,Xue2021} have demonstrated three-dimensional acoustic topological structures with the couplings of non-nearest neighbors, which renders our proposal fully feasible for the state-of-the-art acoustic experiments.

Another promising platform is provided by the resonant electric circuits which allow one to fabricate such three-dimensional topological structures as octupole insulators~\cite{Bao2019,Liu-Ma-Zhang2020} offering an easy access to the sign of the coupling amplitude via the combination of inductive and capacitive impedances. The enormous flexibility in electrical connections enables not only the next-nearest neighbor interactions recently exploited in 2D topolectrical circuits~\cite{Olekhno2021}, but even more exotic physics including four-dimensional quantum Hall effect~\cite{Wang2020} as well as the implementation of a hexadecapole insulator~\cite{Zhang2020}. Thus, electric circuits provide another route to implement our proposal experimentally.


We believe that our results uncover an exciting aspect of topological physics highlighting the role of long-range interactions in the formation of three-dimensional photonic topological phases with possible applications to disorder-robust three-dimensional topological resonators.

\section*{ACKNOWLEDGMENTS}

We acknowledge valuable discussions with Alexander Khanikaev and Nikita Olekhno. This work was supported by the Russian Science Foundation (Grant No.
20-72-10065). M.A.G. acknowledges partial support by the Foundation for the Advancement of Theoretical Physics and Mathematics ``Basis''.

\appendix

\section{Calculation of the topological invariant via symmetry operators eigenvalues}
\label{supp:a}

\begin{figure}[b]
\center{\includegraphics[scale=0.65]{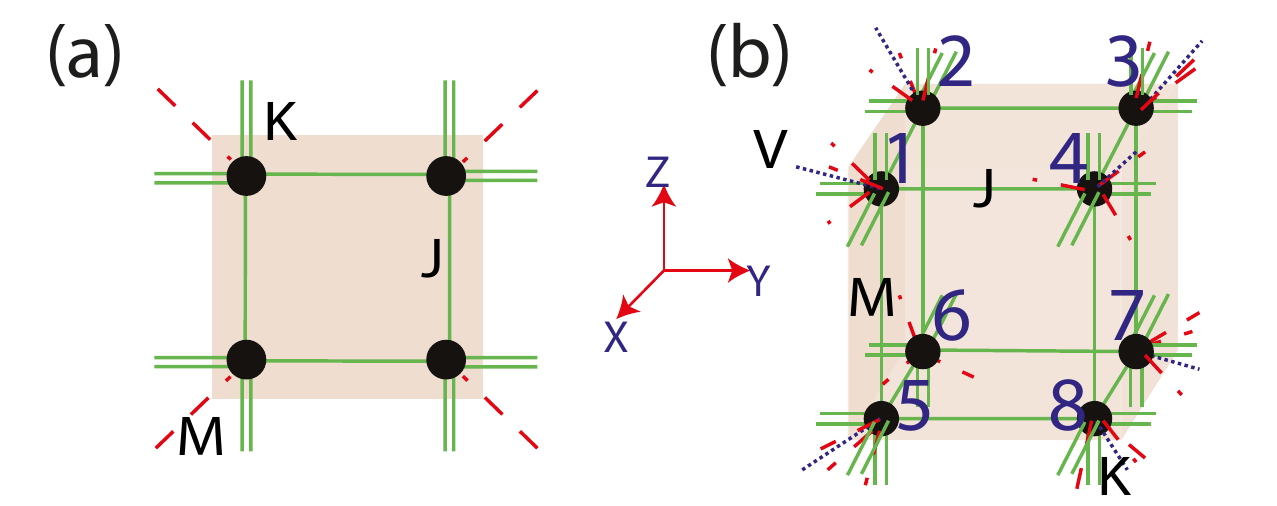}}
\caption{Unit cell geometry for the proposed three-dimensional higher-order topological insulator. Nearest-neighbor couplings $J$ and $K$, both positive, are shown by the single and double green lines, respectively. Two additional types of the next-nearest neighbor coupling with the amplitudes $M$ and $V$ are shown by the red dashed and blue dotted lines, respectively. (a) Single face of the unit cell with the weak links $J$ inside. (b) Three-dimensional unit cell with weak links $J$ inside and enumeration of the sites used to construct Bloch Hamiltonian Eq.~\eqref{aeq:hj}.}\label{fig:wli}
\end{figure}

To calculate the topological invariant, we choose the unit cell with the weak links $J$  inside enumerating the sites of the lattice as shown in Fig.~\ref{fig:wli}. In such case, the Hamiltonian takes the form different from Eq.~\eqref{eq:Hamiltonian}:

\begin{equation}
\begin{gathered}
    H_J(k_x,k_y,k_z)=\left(
        \begin{tabular}{c|c}
          $ H_{J0}(k_x,k_y)$   & $H_{J1}(k_x,k_y,k_z)$  \\
          \hline
          $H_{J1}^\dag(k_x,k_y,k_z)$   & $H_{J0}(k_x,k_y)$
        \end{tabular}
    \right),
\end{gathered}
\label{aeq:hj}
\end{equation}
\begin{widetext}
where the blocks $H_{J0}(k_x,k_y)$ and $H_{J1}(k_x,k_y,k_z)$ read:
\begin{equation}
\label{eq:hamj2}
    H_{J0} = 
    \left( \begin{tabular}{c c c c}
        $0$ & $J + K e^{i k_x}$ &$ M e^{i (k_x-k_y)}$ & $J + K e^{-i k_y}$\\ 
        $J + K e^{-i k_x}$ & $0$ & $J + K e^{-i k_y}$ & $M e^{-i(k_x+k_y)}$ \\ 
        $M e^{-i(k_x-k_y)}$ & $J + K e^{i k_y}$ & $0$ & $J + K e^{-i k_x}$ \\
        $J + K e^{i k_y}$ & $M e^{i(k_x+k_y)}$ & $J + K e^{i k_x}$ & $0$
    \end{tabular}\right),
\end{equation}
\begin{equation}
\label{eq:hamj3}
    H_{J1} = 
    \left( \begin{tabular}{c c c c}
        $J+K e^{i k_z}$ & $M e^{i (k_x+k_z)}$ &$ V e^{i (k_x-k_y+k_z)}$ & $M e^{-i(k_y-k_z)}$\\ 
        $M e^{-i (k_x-k_z)}$ & $J+K e^{i k_z}$ & $M e^{-i (k_y-k_z)}$ & $V e^{-i(k_x+k_y-k_z)}$ \\ 
        $V e^{i (-k_x+k_y+k_z)}$ & $M e^{i(k_y+k_z)}$ & $J+K e^{i k_z}$ & $M e^{i(-k_x+k_z)}$ \\
        $M e^{i(k_y+k_z)}$ & $V e^{i (k_x+k_y+k_z)}$ & $M e^{i(k_x+k_z)}$ & $J+K e^{i k_z}$
    \end{tabular}\right),
\end{equation}
\end{widetext}

Investigating the topological properties of the system, we use the fact that the Hamiltonian commutes with the symmetry operators in certain high-symmetry points of the first Brillouin zone: $H_J$ commutes with $i$ and $C_2$ in $X$, $M$ and $R$ points and commutes with $C_3$ and $C_4$ operators only in $R$ point. Since the associated topological indices are not fully independent, we choose the set of numbers given by Eq.~\eqref{eq:chi}.

To extract the topological invariant, we examine the behavior of the eigenstates in the high-symmetry points mentioned above under the action of the respective symmetry operators. We start from $C_4$ symmetry operator describing the rotation around $z$ axis by $\pi/2$ angle, which has the following form:
\begin{equation}
    \widehat{C}_4=\left(
        \begin{tabular}{c|c}
          $C_4$ & $0_4$\\
          \hline
          $0_4$ & $C_4$
        \end{tabular}
    \right),
\end{equation}
where a $4\times 4$ block $C_4$ reads:
\begin{equation}
    C_4=\left(
        \begin{tabular}{c c c c}
        0 & 1 & 0 & 0 \\
        0 & 0 & 1 & 0 \\
        0 & 0 & 0 & 1 \\
        1 & 0 & 0 & 0 \\
        \end{tabular}
    \right)
\end{equation}
and $0_4$ is a $4\times 4$ zero matrix.

Rotations by the angle $\pi$ around $z$ axis are described by $\widehat{C}_2$ operator, which takes the form:
\begin{equation}
    \widehat{C}_2=\left(
        \begin{tabular}{c|c}
          $C_2$& $0_4$\\
          \hline
          $0_4$ & $C_2$
        \end{tabular}
    \right),
\end{equation}
where $4\times 4$ block $C_2$ is given by
\begin{equation}
    C_2=\left(
        \begin{tabular}{c c c c}
        0 & 0 & 1 & 0 \\
        0 & 0 & 0 & 1 \\
        1 & 0 & 0 & 0 \\
        0 & 1 & 0 & 0 \\
        \end{tabular}
    \right).
\end{equation}
Spatial inversion $\widehat{i}$ is described by the operator
\begin{equation}
    \widehat{i}=\left(
        \begin{tabular}{c|c}
         $0_4$ & $C_2$ \\
          \hline
           $C_2$ & $0_4$
        \end{tabular}
    \right).
\end{equation}
Finally, rotations by the angle $2\pi/3$ with respect to the axis defined by  $-\bas{x}+\bas{y}+\bas{z}$ vector are described by the matrix
\begin{equation}
    \widehat{C}_3=\left(
        \begin{tabular}{c c c c c c c c}
        0 & 0 & 0 & 0 & 0 & 0 & 0 & 1 \\
        0 & 0 & 0 & 1 & 0 & 0 & 0 & 0 \\
        0 & 0 & 1 & 0 & 0 & 0 & 0 & 0 \\
        0 & 0 & 0 & 0 & 0 & 0 & 1 & 0 \\
        0 & 0 & 0 & 0 & 1 & 0 & 0 & 0 \\
        1 & 0 & 0 & 0 & 0 & 0 & 0 & 0 \\
        0 & 1 & 0 & 0 & 0 & 0 & 0 & 0 \\
        0 & 0 & 0 & 0 & 0 & 1 & 0 & 0 \\
        \end{tabular}
    \right).
\end{equation}

Once symmetry operators are specified, we can assess the behavior of all eigenstates with respect to these symmetry transformations  for the wave vectors $(k_x,k_y, k_z) = (0,0,0)$, $(\pi,0,0)$, $(\pi,\pi,0)$ and $(\pi,\pi,\pi)$ in analogy to the two-dimensional case~\cite{Benalcazar2019}.  The results calculated for the fixed parameters $J= 1$, $K= 6$, $M=4$ and $V=-3$ are summarized in Table~\ref{atab:Topological}.

\begin{small}
\begin{table*}
\begin{center}
\begin{tabular}{l| l l c| l l c}
\toprule
$i$ & \multicolumn{3}{l|} {$\Gamma$=($k_x=k_y=k_z=0$)} & \multicolumn{3}{l} {$X$=($k_x=\pi,k_y=k_z=0$)} \\
\toprule
 & $\centering{\varepsilon} $  & $\centering{\psi} $ & $\centering{p(i)}$ & $\centering{\varepsilon} $  & $\centering{\psi} $ & $\centering{p(i)}$ \\
 \cmidrule(r){2-7}
 & $\varepsilon_1=30$  &  $\psi_1=(1,1,1,1,1,1,1,1)$ & $1$ & $\varepsilon_1=28$  &  $\psi_1=(1,-1,-1,1,1,-1,-1,1)$ & $2$\\
 & $\varepsilon_2=6 $  &  $\psi_2=(0,-1,-1,0,1,0,0,1)$ & $2$ & $\varepsilon_1=8$  &  $\psi_2=(1,1,1,1,1,1,1,1)$ & $1$\\
 & $\varepsilon_3=6 $  &  $\psi_3=(-1,0,1,0,-1,0,1,0)$ & $2$ & $\varepsilon_1=4$  &  $\psi_3=(-1, 1, 0, 0, 0, 0, -1, 1)$ & $1$\\
 & $\varepsilon_4=6 $  &  $\psi_4=(0,0,-1,-1,1,1,0,0)$ & $2$ & $\varepsilon_1=4$  &  $\psi_4=(0, 0, -1, 1, -1, 1, 0, 0)$ & $1$\\
 \toprule
$C_2$ & \multicolumn{3}{l|} {$\Gamma$=($k_x=k_y=k_z=0$)} & \multicolumn{3}{l} {$M$=($k_x=k_y=\pi,k_z=0$)} \\
 \toprule
  & $\centering{\varepsilon} $  & $\centering{\psi} $ & $\centering{p(C_{2})}$ & $\centering{\varepsilon} $  & $\centering{\psi} $ & $\centering{p(C_{2})}$ \\
 \cmidrule(r){2-7}
 & $\varepsilon_1=30$  &  $\psi_1=(1,1,1,1,1,1,1,1)$ & $1$ & $\varepsilon_1=26$  &  $\psi_1=(-1, 1, -1, 1, -1, 1, -1, 1)$ & $1$\\
 & $\varepsilon_2=6 $  &  $\psi_2=(0,-1,1,2,-1,-2,0,1)$ & $2$ & $\varepsilon_1=6$  &  $\psi_2=(0, -1, 0, 1, 0, -1, 0, 1)$ & $2$\\
 & $\varepsilon_3=6 $  &  $\psi_3=(-1,0,1,0,-1,0,1,0)$ & $2$ & $\varepsilon_1=6$  &  $\psi_3=(-1, 0, 1, 0, -1, 0, 1, 0)$ & $2$\\
 & $\varepsilon_4=6 $  &  $\psi_4=(0,-1,-3,-2,3,2,0,1)$ & $1$ & $\varepsilon_1=2$  &  $\psi_4=(1,-1,1,-1,-1,1,-1,1)$ & $1$\\
 \toprule
$C_3$ & \multicolumn{3}{l|} {$\Gamma$=($k_x=k_y=k_z=0$)} & \multicolumn{3}{l} {$R$=($k_x=k_y=k_z=\pi$)} \\
 \toprule
  & $\centering{\varepsilon} $  & $\centering{\psi} $ &  $\centering{p(C_{3})}$ & $\centering{\varepsilon} $  & $\centering{\psi} $ &  $\centering{p(C_{3})}$\\
 \cmidrule(r){2-7}
 & $\varepsilon_1=30$  &  $\psi_1=(1,1,1,1,1,1,1,1)$ & $1$ & $\varepsilon_1=24$  &  $\psi_1=(1, -1, 1, -1, -1, 1, -1, 1)$ & $1$\\
 & $\varepsilon_2=6 $  &  $\psi_2=(1,-e^{2i\pi/3}, 0, e^{i\pi/3}, 0, -e^{i\pi/3}, -1, e^{2i\pi/3})$ & $2$ & $\varepsilon_1=4$  &  $\psi_2=(1, e^{2i\pi/3}, 0, -e^{i\pi/3}, 0, -e^{i\pi/3}, 1, e^{2i\pi/3})$ & $2$\\
 & $\varepsilon_3=6 $  &  $\psi_3=(-e^{i\pi/3}, -e^{2i\pi/3}, 0, -1, 0, 1, e^{i\pi/3}, e^{2i\pi/3})$ & $3$ & $\varepsilon_1=4$  &  $\psi_3=(1, -e^{i\pi/3}, 0, e^{2i\pi/3}, 0, e^{2i\pi/3}, 1, -e^{i\pi/3})$ & $3$\\
 & $\varepsilon_4=6 $  &  $\psi_4=(1, 1, -3, 1, -3, 1, 1, 1)$ & $1$ & $\varepsilon_1=4$  &  $\psi_4=(1, -1, -3, -1, 3, 1, -1, 1)$ & $1$\\
  \toprule
$C_4$ & \multicolumn{3}{l|} {$\Gamma$=($k_x=k_y=k_z=0$)} & \multicolumn{3}{l} {$R$=($k_x=k_y=k_z=\pi$)} \\
 \toprule
  & $\centering{\varepsilon} $  & $\centering{\psi} $ &  $\centering{p(C_{4})}$ & $\centering{\varepsilon} $  & $\centering{\psi} $ &  $\centering{p(C_{4})}$ \\
\cmidrule(r){2-7}
 & $\varepsilon_1=30$  &  $\psi_1=(1,1,1,1,1,1,1,1)$ & $1$ & $\varepsilon_1=24$  &  $\psi_1=(1, -1, 1, -1, -1, 1, -1, 1)$ & $3$\\
 & $\varepsilon_2=6 $  &  $\psi_2=(-i, 1, i, -1, -i, 1, i, -1)$ & $2$ & $\varepsilon_1=4$  &  $\psi_2=(i, 1, -i, -1, -i, -1, i, 1)$ & $4$\\
 & $\varepsilon_3=6 $  &  $\psi_3=(-i, -1, i, 1, -i, -1, i, 1)$ & $4$ & $\varepsilon_1=4$  &  $\psi_3=(i, -1, -i, 1, -i, 1, i, -1)$ & $2$\\
 & $\varepsilon_4=6 $  &  $\psi_4=(-1, -1, -1, -1, 1, 1, 1, 1)$ & $1$ & $\varepsilon_1=4$  &  $\psi_4=(-1, 1, -1, 1, -1, 1, -1, 1)$ & $3$\\
 \hline
\end{tabular}
\caption{Calculation of the topological invariant for the unit cell with the weak links inside. Parameters of the Hamiltonian read: $J= 1$, $K= 6$, $M=4$ and $V=-3$. $\varepsilon_i$ and $\psi_i$ are the first four eigenvalues above the bandgap and the corresponding eigenfunctions calculated for the Bloch Hamiltonian $H_J(k_x,k_z,k_y)$, Eq.~(\ref{aeq:hj}) for $i=1...4$. $p(C_2)$ and $p(i)$ are indices of eigenvalues of the  rotation operator $\widehat{C}_2$ and inversion $\widehat{i}$ given by the expressions $\exp(i\pi(p(C_2)-1))$ and $\exp(i\pi(p(i)-1))$, respectively. $p(C_3)$ and $p(C_4)$ are indices of eigenvalues of the  rotation operators $\widehat{C}_3$ and $\widehat{C}_4$ given by the expressions $\exp(2\pi\,i(p(C_3)-1)/3)$ and $\exp(i\pi(p(C_4)-1)/2)$, respectively.\label{atab:Topological}}
\end{center}
\end{table*}
\end{small}

Using Table~\ref{atab:Topological} and checking the symmetry of the states only above zero-energy bandgap, we recover:
\begin{align*}
    \tilde{\#}{\rm X}_{1}^{(i)}&=3, \quad \tilde{\#}\Gamma_{1}^{(i)} = 1,\\
    \tilde{\#}{\rm M}_{1}^{(C_2)}&=2, \quad \tilde{\#}\Gamma_{1}^{(C_2)} = 2,\\
    \tilde{\#}{\rm R}_{1}^{(C_3)}&=2, \quad \tilde{\#}\Gamma_{1}^{(C_3)} = 2,\\
    \tilde{\#}{\rm R}_{1}^{(C_4)}&=0, \quad \tilde{\#}\Gamma_{1}^{(C_4)} = 2,   
\end{align*}
where $\tilde{\#}{\rm \Pi}$ stands for the number of states with a given symmetry above the bandgap. Clearly, it is related to the number of states below the band gap $\#{\rm \Pi}$ as $\tilde{\#}{\rm \Pi}=8-\#{\rm \Pi}$ (the total number of bands is equal to 8), and hence  Eq.~(\ref{eq:chi}) yields the topological invariant $\chi=(-2,0,0,2)$ which indicates the topological case. This result aligns with our calculation for the finite structure in Sec.~\ref{sec:finite}, which demonstrates the corner state at the weak link corner.

\section{Bulk polarization: symmetry constraints and evaluation from the symmetry indices}
\label{supp:b}

In this Appendix, we discuss the constraints imposed by the cubic symmetry of the lattice on the values of bulk polarization. We define the translation vectors of the cubic lattice as $\textbf{a}_1=(1,0,0)$, $\textbf{a}_2=(0,1,0)$, $\textbf{a}_3=(0,0,1)$ and expand bulk polarization in terms of them: $\textbf{P}=\sum_{i=1}^3 p_i\textbf{a}_i$. 

Transformation described by the matrix $\hat{T}$ transforms the lattice vectors as  $\textbf{a}_i' = \sum_{j=1}^3 T_{ij}\textbf{a}_j$ and the polarization transforms as
\begin{equation}
    \textbf{P}=\sum_{i=1}^3 p_i\textbf{a}_i \rightarrow \sum_{i,j=1}^3 p_i\,T_{ij}\textbf{a}_j.
    \label{apeq:C1}
\end{equation}

If $\hat{T}$ is the symmetry transformation, bulk polarization of the structure Eq.~\eqref{apeq:C1} must not change. Given that the polarization is defined up to the translation vector, the following transformation law is fulfilled:
\begin{equation}
    \textbf{P}=\sum_{i=1}^3 p_i\textbf{a}_i \rightarrow \sum_{i=1}^3 (p_i+n_i)\textbf{a}_i,
    \label{apeq:C2}
\end{equation}
where $n_i \in \mathbb{Z}$, $i=1,2,3$. Comparing Eq.~\eqref{apeq:C1} and Eq.~\eqref{apeq:C2}, we derive the constraints on the possible values of bulk polarization: 
\begin{equation}
    \sum_{i=1}^3 p_i T_{ij} = (p_j+n_j),
    \label{apeq:C3}
\end{equation}
where integers $n_j$ depend on the chosen transformation $\hat{T}$.

As symmetry transformations, $O_h$ symmetry group includes $\pi/2$ rotations $\textbf{T}_z$ and $\textbf{T}_y$ with respect to $z$ and $y$ axes: 
\begin{equation}
    \textbf{T}_z=\left(
        \begin{tabular}{c c c}
        0 & -1 & 0  \\
        1 & 0 & 0  \\
        0 & 0 & 1  \\
        \end{tabular}
    \right),
    \label{apeq:C4}
\end{equation}
\begin{equation}
    \textbf{T}_y=\left(
        \begin{tabular}{c c c}
        0 & 0 & 1  \\
        0 & 1 & 0  \\
        -1 & 0 & 0  \\
        \end{tabular}
    \right).
    \label{apeq:C5}
\end{equation}

Plugging the matrices  $\textbf{T}_z$ and $\textbf{T}_y$ in Eq.~\eqref{apeq:C3}, we derive the polarization components $p_1$, $p_2$, $p_3$:
\begin{equation}
    \begin{split}
    \text{for $\textbf{T}_z$} \\ &  p_1 = -\frac{n_1+n_2}{2},\quad p_2=\frac{n_1-n_2}{2}; \\
     \text{for $\textbf{T}_y$}\\ & p_1 = \frac{-\tilde{n}_1+\tilde{n}_3}{2}, \quad p_3=-\frac{\tilde{n}_1+\tilde{n}_3}{2}.\\
    \end{split}
\end{equation}
Since all components of bulk polarization are defined modulo 1, $p_1$, $p_2$ and $p_3$ are quantized to be either $0$ or $1/2$. Furthermore, $p_1-p_2=-n_1$ and $p_1-p_3=\tilde{n}_3$ and hence all components of bulk polarization are equal to each other modulo 1.

Thus, there are two possibilities for the bulk polarization in $O_h$-symmetric system: either trivial $(0,0,0)$ or non-trivial $(1/2,1/2,1/2)$.

As we show below, the choice of the particular option is governed by the inversion symmetry of eigenstates in $\Gamma$ and $X$ points of the Brillouin zone. To this end, we explicitly write bulk polarization in terms of occupied states $\ket{u_{\bf k}^m}$~\cite{KingSmith1993}:
\begin{equation}
    \textbf{P}=\frac{1}{(2\pi)^3}\int_{BZ} \sum_m \expval{ \frac{\partial}{\partial {\bf k}}}{u^m_{\bf k}} d^3{\bf k}.
    \label{apeq:pol}
\end{equation}
Next we consider the inversion $J$ which satisfies the condition $\widehat{J}\,  ^2=\widehat{I}$ ($\widehat{I}$ is the identity matrix) and transforms the Hamiltonian as $\widehat{J}\, \widehat{h}(\vec{k})\widehat{J}^{-1}= \widehat{h}(-\vec{k})$. Introducing the eigenvector $\ket{u_{\bf k}^l}$ corresponding to the band with the index $l$, we recover
\begin{equation}\label{eq:SewingM}
    \begin{gathered}
        \varepsilon_{l {\bf k}} \widehat{J} \ket{u_{{\bf k}}^l}=\widehat{J}\,\widehat{h}(\vec{k}) \ket{u_{{\bf k}}^l} = \widehat{J}\,\widehat{h}({\bf k})\,\widehat{J}\,\widehat{J}\, \ket{u_{{\bf k}}^l} 
        =\widehat{h}(-{\bf k})\widehat{J}\ket{u_{{\bf k}}^l}\:,
    \end{gathered}
\end{equation}
where $\widehat{h}(\vec{k})$ is Bloch Hamiltonian of the system and $\ket{u_{{\bf k}}^l}$ is the eigenstate with energy $\varepsilon_{l \vec{k}}$. Equation~\eqref{eq:SewingM} thus shows that $\widehat{J}\ket{u_{{\bf k}}^l}$ is an eigenstate of the Hamiltonian $\widehat{h}(-{\bf k})$ with the energy $\eps_{l{\bf k}}$. Hence, this eigenstate can be expanded in terms of $\widehat{h}(-{\bf k})$ eigenstates:
\begin{equation}\label{eq:SewingM2}
\widehat{J}\ket{u_{{\bf k}}^l}=\sum_m \ket{u_{-{\bf k}}^m}B_{{\bf k}}^{ml}\:,   
\end{equation}
where $\hat{B}_{\bf k}$ is the inversion sewing matrix. Due to orthogonality of eigenstates, the components of this matrix read:
\begin{equation}
\begin{gathered}
    B_{{\bf k}}^{ml} = \bra{u_{-{\bf k}}^m} \widehat{J}\ket{u_{{\bf k}}^l},\\
\end{gathered}
\end{equation}
and it is straightforward to verify that the sewing matrix is unitary. In a suitable basis, this matrix can be brought to the diagonal form. In that case, the entries at its diagonal provide the inversion parity of the eigenstates, being either $1$ or $e^{i\,\pi}$. Hence, the logarithm of the determinant can be presented as
\begin{equation}
    \log \, \text{det} B_{{\bf k}} = i\pi \,\#\Pi_2^{J},
\end{equation}
where point $\Pi$ in the Brillouin zone corresponds to the wave vector ${\bf k}$ and $\#\Pi_2^{J}$ denotes the number of occupied eigenstates which are odd under inversion.

Using these properties and denoting $k_n=2\pi/a\,s_n$, we insert $\widehat{J}^2=\widehat{I}$ in the definition of bulk polarization and derive:
\begin{equation}
\begin{gathered}
    P_i=\frac{1}{2 \pi} \int_{-\frac12}^{\frac12}\int_{-\frac12}^{\frac12}\int_{-\frac12}^{\frac12} \sum_l i \expval{\frac{\partial}{\partial s_i}}{u^l(\vec{s}\,)} d^3{\bf s}=\\
    =\frac{1}{2 \pi} \int_{-\frac12}^{\frac12}\int_{-\frac12}^{\frac12}\int_{-\frac12}^{\frac12} \sum_{l,m,m'} i \bra{u^m(-\vec{s}\,)B_{\vec{s}}^{ml}} \\
    \qquad \frac{\partial}{\partial s_i} \ket{B_{\vec{s}}^{m'l} u^{m'}(-\vec{s})}d^3\vec{s}=\\
    =\frac{1}{2 \pi} \int_{-\frac12}^{\frac12}\int_{-\frac12}^{\frac12}\int_{-\frac12}^{\frac12} \sum_m i\,\expval{\frac{\partial}{\partial s_i}}{u^m(-\vec{s})} d^3\vec{s}+\\
    +\frac{i}{2 \pi} \int_{-\frac12}^{\frac12}\int_{-\frac12}^{\frac12}\int_{-\frac12}^{\frac12} \sum_{l,m,m'} (B_{\vec{s}}^{ml})^* \frac{\partial B_{\vec{s}}^{m'l}}{\partial s_i} d^3 \vec{s}=\\
    =-P_i+\frac{i}{2 \pi} \int_{-\frac12}^{\frac12}\int_{-\frac12}^{\frac12}\int_{-\frac12}^{\frac12}  \text{Tr}\left[ \frac{\partial B_{\vec{s}}}{\partial s_i} B_{\vec{s}}^\dagger \right]\,d^3 \vec{s}=\\
    =-P_i+\frac{i}{2 \pi} \int_{-\frac12}^{\frac12}\int_{-\frac12}^{\frac12}\int_{-\frac12}^{\frac12} \frac{\partial}{\partial s_i} \left( \log\, \text{det}\,B_{\vec{s}} \right)\,d^3\vec{s}.
\end{gathered}
\end{equation}
Here, in lines 2 and 3 we use Eq.~\eqref{eq:SewingM2}, in lines 4 and 5~-- the fact that the matrix $B_{{\bf s}}$ is unitary, in line 7~-- the property $\text{Tr}\,\log\,\hat{M}=\log\,\det\,\hat{M}$.

The obtained result can be simplified further given that the Chern number is zero and hence a smooth gauge for the eigenstates can be chosen. Therefore, the expression under the integral can be made independent of all $s_n$ except $s_i$ by the proper gauge choice~\cite{Benalcazar2019}. With this simplification, bulk polarization component $P_1$ reads:
\begin{equation}
\begin{gathered}
    P_1=\frac{i}{4 \pi} \int_{-\frac12}^{\frac12} \frac{\partial}{\partial s_i} \left( \log\, \text{det}\, B(s_1,0,0) \right)\,ds_1=\\
    = \frac{i}{2 \pi} \int_{0}^{\frac12} \frac{\partial}{\partial s_i} \left( \log\, \text{det}\,B(s_1,0,0) \right)\,ds_1=\\
    =\frac{i}{2 \pi} \left(i\pi \# X_2^{\widehat{J}}-i\pi \# \Gamma_2^{\widehat{J}}  \right)= -\frac{1}{2} \left[ X_2^{J} \right]\:.
\end{gathered}
\end{equation}
In the second line we took into account that $B_{-{\bf s}}=B^\dag_{\bf s}=B^{-1}_{\bf s}$, i.e. $\log\,\det\,B_{-{\bf s}}=-\log\,\det\,B_{\bf s}$. Since bulk polarization is defined modulo 1, it can be also presented in the form
\begin{equation}\label{apeq:finpol}
    P_1=\frac{1}{2} \left[ X_2^{J} \right]\:.
\end{equation}
In our case, $\left[ X_2^{J} \right]$ is either $-2$, or 0 depending on the unit cell choice. Thus, Eq.~\eqref{apeq:finpol}  yields that bulk polarization for the four bands below  zero-energy bandgap is equal to zero for both unit cell choices.

\section{Evaluation of the corner charge}\label{supp:c}

\begin{figure}[b!]
\center{\includegraphics[scale=0.5]{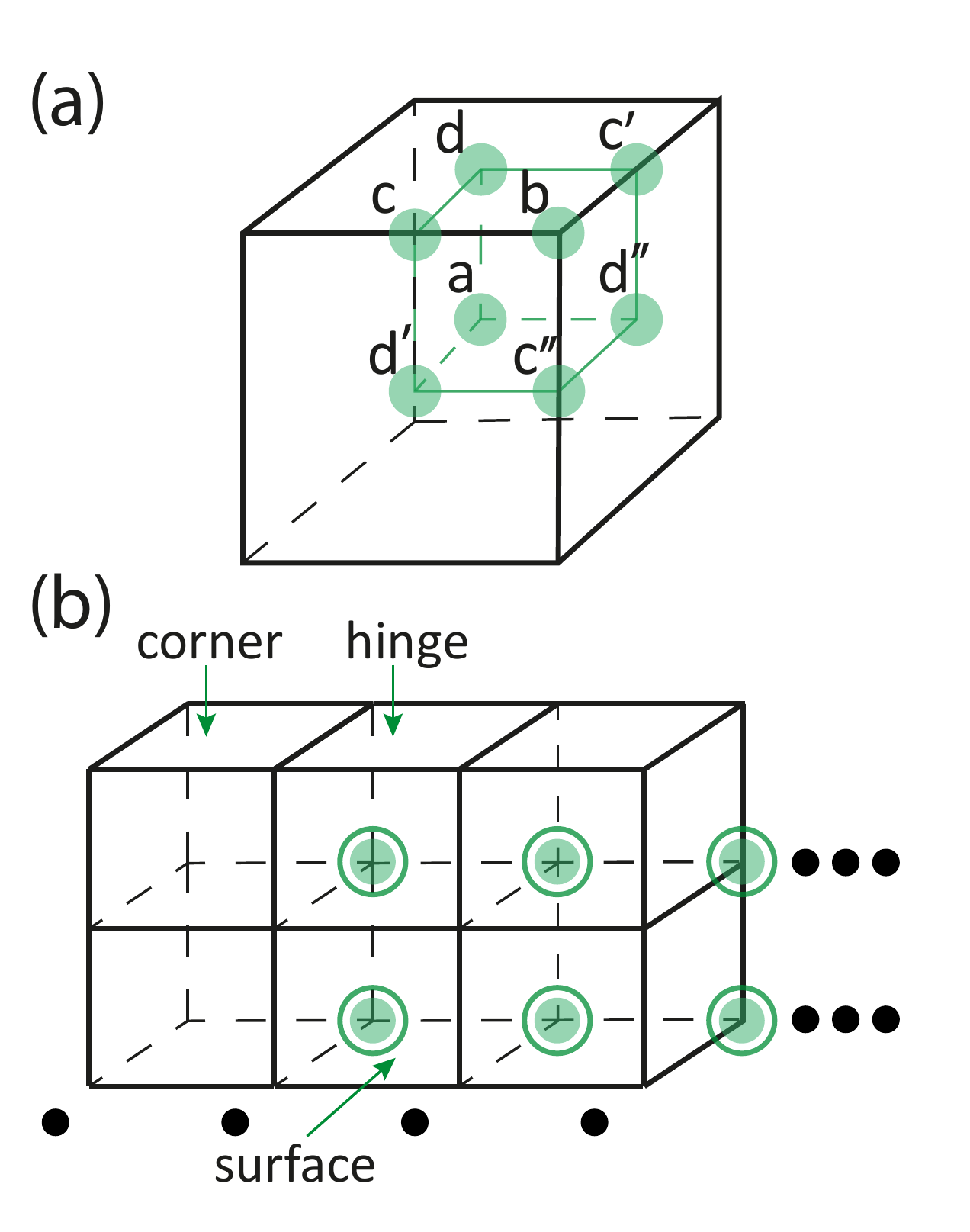}}
\caption{Illustration of corner charge calculation. (a) Maximal Wyckoff positions in the cubic unit cell possessing $O_h$ symmetry. Due to symmetry, Wannier centers can appear only in the specified points.  (b) Semi-infinite sample of the considered structure with $O_h$ symmetry and illustration of the corner charge calculation.}\label{afig:Wannier}
\end{figure}

In this Appendix, we consider the unit cell with the weak links inside that corresponds to the topological scenario. While bulk polarization of the four bands below zero-energy bandgap vanishes, two out of four occupied bands have nontrivial polarization $(1/2,1/2,1/2)$. This means that two Wannier centers occupy {\it a} position in the unit cell, while the other two are located in position {\it b} [Fig.~\ref{afig:Wannier}(a)].

Note that symmetry allows also other positions of Wannier centers: $c$, $c'$, $c''$ or $d$, $d'$, $d''$. However, this option requires at least three bands with the nontrivial polarization which is not the case for our system.

Next we consider the structure with a finite number of unit cells, each has an associated pair of Wannier centers at the corner (i.e. at position {\it b}). Since the distribution of the  charge must be consistent with the symmetry of the system, the neutrality of the system breaks down and some of Wannier centers should be removed (so-called filling anomaly). Unit cell in the bulk has 16 Wannier centers at the corners which contribute charge equal to  $2$. Unit cell at the surface has only 8 Wannier centers receiving a share of charge equal to $1$. Since the charge is defined modulo 1, these cells remain neutral. However, corner unit cell has only 2 Wannier centers at one corner which contribute charge $1/4$. Hence, we conclude that all weak-link corners of the designed structure have a quantized corner charge equal to $1/4$.

\section{Local density of states}
\label{supp:d}

\begin{figure}[b]
\center{\includegraphics[scale=0.7]{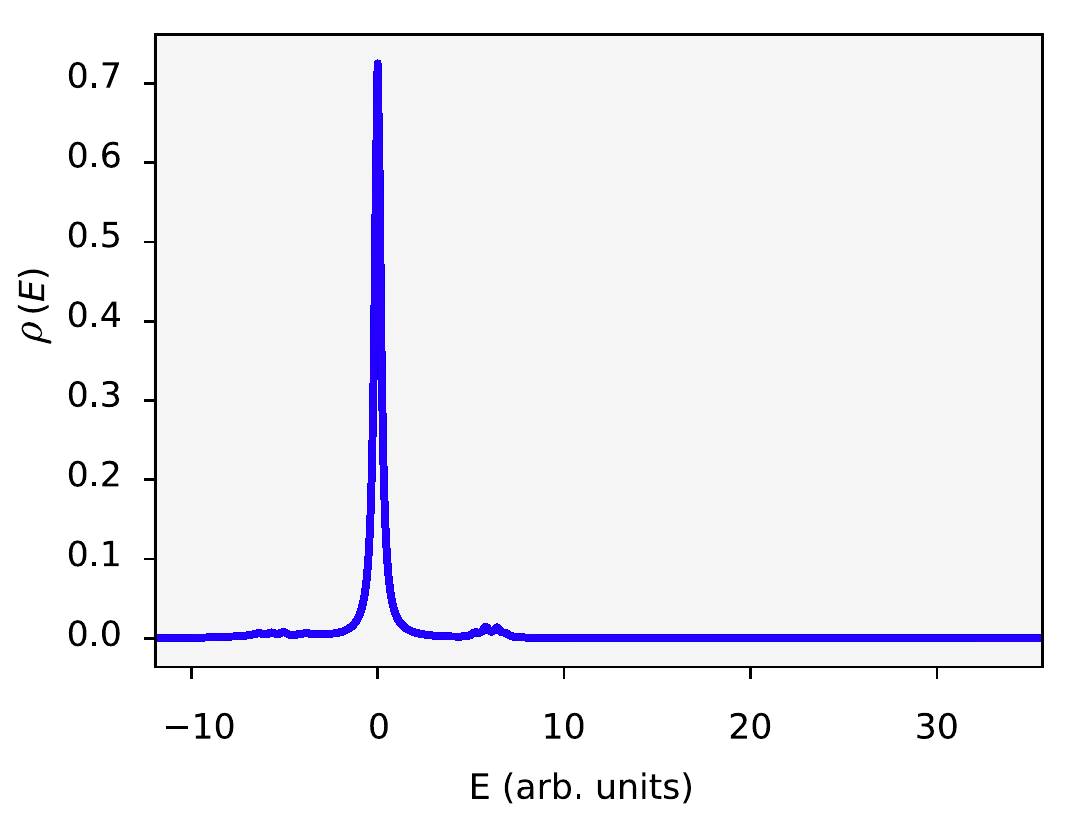}}
\caption{Local density of states calculated at the weak link corner of a $9\times 9 \times 9$ system with $J= 1$, $K= 6$, $M=4$ and $V=3$, when the corner mode appears in the continuum. Auxiliary parameter $\gamma$ responsible for the width of the peak is set to $0.2$.}
\label{fig:ldos}
\end{figure}

In this Appendix, we discuss how to probe the topological corner state if it arises inside the continuum of the bulk modes. A key feature that allows one to discriminate the bound state from the bulk modes is good spatial localization of the bound mode at the corner of the lattice. Consider the local density of states (LDOS) defined as 
\begin{equation}\label{eq:ldos}
    \rho(E) = 
    \sum_a \frac{\gamma^2}{\gamma^2+(E-\varepsilon (a))^2}|\Psi_{nnn}(a)|^2,
\end{equation}
where $\varepsilon(a)$ is the energy of the eigenmode with number $a$ and $\Psi_{nnn}(a)$ is the amplitude of the respective wave function at the weak link corner of the lattice. In this calculation, the squared amplitude of the wave function is multiplied by the Lorentzian weighting factor which has a maximum at energy $E$ corresponding to the energy of the respective eigenmode, $\varepsilon (a)$. Auxiliary parameter $\gamma$ controls the width of the associated peak.

Local density of states calculated according to Eq.~\eqref{eq:ldos} for BIC regime is depicted in Fig.~\ref{fig:ldos}. Even though corner state spectrally overlaps with the bulk modes, calculated $\rho(E)$ function features a clear peak proving the existence of the corner-localized mode.

Importantly, local density of states can be directly probed in experiments since it is linked to the radiation resistance of an antenna exciting the structure. Therefore, measurement of LDOS function allows one to access BIC states in experiments as has been  recently done for 2D topological structure~\cite{Peterson1114}.


\section{Interface states and disorder-robustness}\label{supp:e}

\begin{figure}[t!]
\center{\includegraphics[scale=0.65]{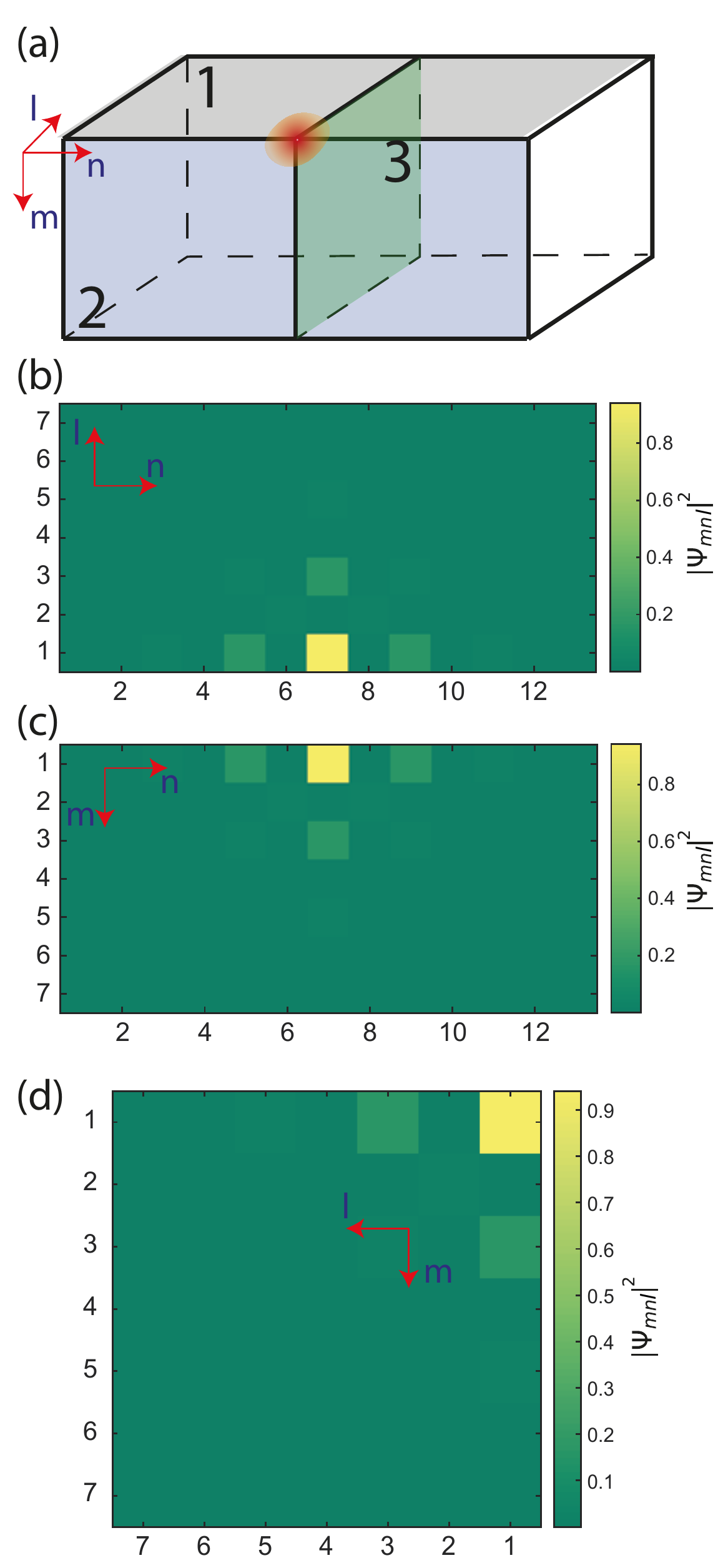}}
\caption{Localization of the topological corner mode at the interface between the two domains with the different dimerizations. (a) The sketch of the considered structure. Red dot highlights the position of the topological interface mode. (b-d) Eigenmode profile at the faces No. 1-3 calculated for $J=1$, $K=6$, $M=4$, $V=-3$.}\label{afig:inter}
\end{figure}

The topological origin of our system is manifested not only in the corner-localized states, but also in the zero-dimensional interface states which arise at the boundary of the two structures with the different dimerizations. To illustrate this, we consider $7\times 7\times 13$ structure which consists of the two blocks with different dimerizations sharing the common boundary and a weak link corner [Fig.~\ref{afig:inter}(a)]. Calculating the eigenmodes of this structure, we recover a zero-dimensional state localized at the interface and having the energy close to that of the corner state. This indicates that the topological properties of our model are revealed in the variety of ways depending on the sample geometry.

\begin{figure}[t!]
\center{\includegraphics[scale=0.8]{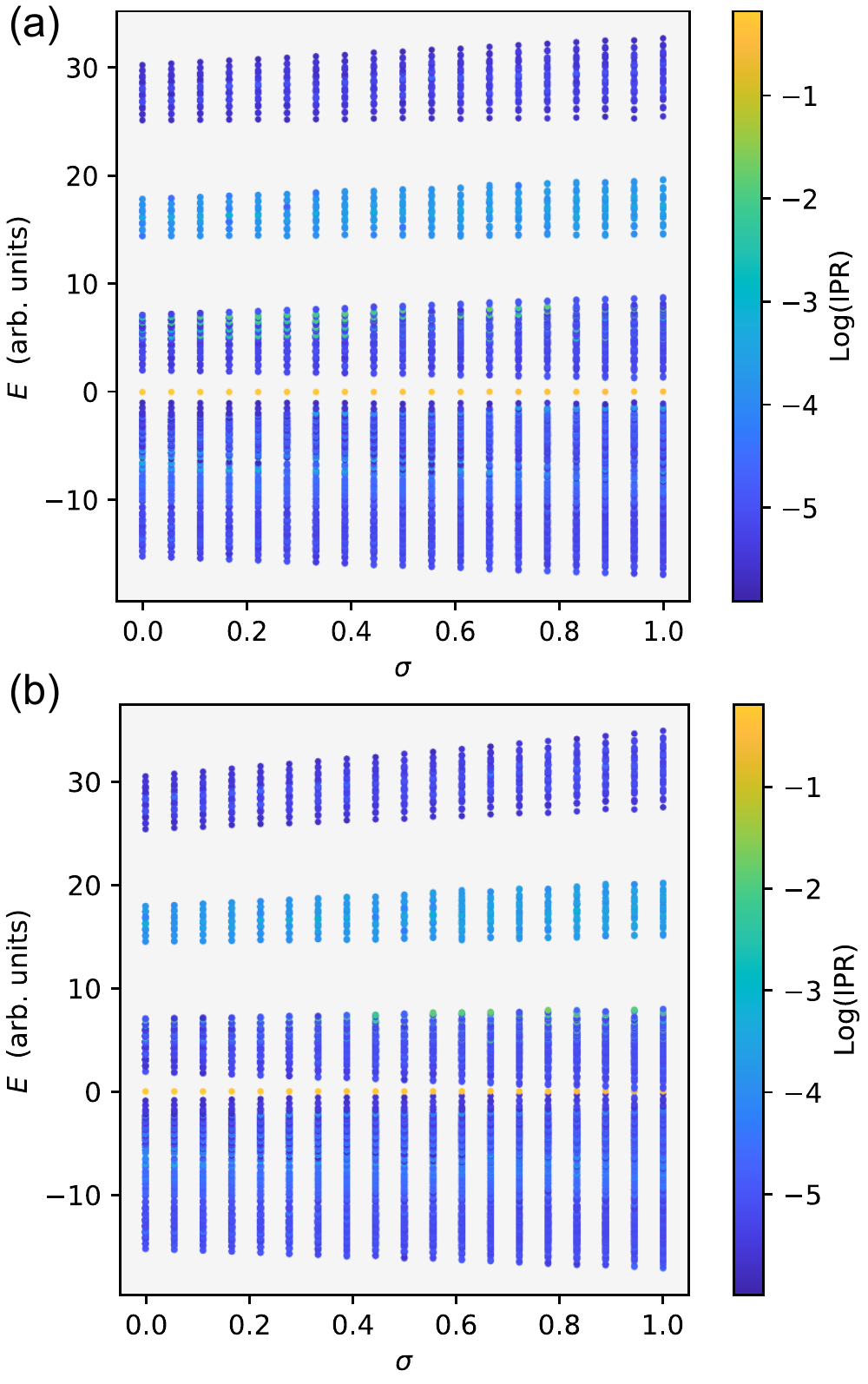}}
\caption{Eigenmodes of a finite 3D $9 \times 9 \times 9$ structure versus disorder strength $\sigma$ averaged over 10 random realizations. The disorder is introduced in the coupling constants as follows: (a) $J=J_0+\sigma \Delta J$, $K=K_0+\sigma \Delta K$; (b) $J=J_0+\sigma \Delta J$, $K=K_0+\sigma \Delta K$, $M=M_0+\sigma \Delta M$, $V=V_0+\sigma \Delta V$, where coupling constants $J_0 = 1$, $K_0 = 6$, $M_0= 4$, $V_0=-2.5$. $\Delta K$, $\Delta J$, $\Delta M$, $\Delta V$ are uniformly distributed random numbers in the interval [0,1]. Color shows the logarithm of the inverse participation ratio. Orange dots correspond to the corner state.}\label{afig:robust}
\end{figure}

Another important aspect is the robustness of predicted corner states to the various types of disorder. Similarly to the paradigmatic Su-Schrieffer-Heeger model, the corner states are quite robust to the coupling disorder remaining unprotected from on-site disorder. Therefore, we analyze two scenarios: disorder in the nearest-neighbor couplings and disorder in all coupling links [Fig.~\ref{afig:robust}(a,b), respectively].

In both cases, we observe the narrowing of the bandgap with the disorder strength. However, the corner state persists retaining relatively good localization as quantified by the inverse participation ratio. Additionally, the corner state features greater robustness to the variation of the nearest-neighbor coupling compared to the long-range one.

\bibliography{TopologicalLib}

\end{document}